# Correlation between measured and unmeasured quantities in quantum mechanics, locality, and the legitimacy of counterfactual reasoning

Oliver Cohen

14 Nightingale Close

Winchester

SO22 5QA

UK

oliver.cohen@hotmail.co.uk

**A correlation measure relating to measured and unmeasured local quantities in quantum mechanics is introduced, and is then applied to assess the locality implications for Bell/CHSH and similar set-ups. This leads to some interesting results, and the scheme is extended to the generalized no-signalling boxes framework. Some questions are raised about the use of counterfactual reasoning in quantum mechanics.**

As we get closer to the fiftieth anniversary of Bell's derivation of his inequality [1], discussions about the validity and significance of the inequality, and what it reveals about the meaning and structure of quantum mechanics, continue to roll on. Whilst the current majority view seems to be that a Bell inequality violation can be understood as proof of an inescapable nonlocality in nature, there still exist a good number of vociferous dissenters to this viewpoint. Recently, the new concept of Information Causality has provided valuable insight into why the maximal violation of the Bell/ CHSH inequality allowed by quantum mechanics (the Tsierlson bound) is what it is [2]. Quite separately from this, there has been some progress [3] in clarifying the role of counterfactual reasoning in Bell's derivation and in the derivation of CHSH's [4] subsequent more general formulation of the inequality . And in another independent development [5], the notion of a symmetric informationally complete positive-operator-valued measure ("SIC") has been proposed, in a Bayesian framework, as, among other things, a means of providing insight into how Bell's theorem should be interpreted.

In the current paper, a new measure is introduced to assess the correlation between locally measured and unmeasured quantities, and the measure is then applied to a Bell/CHSH set-up. It transpires that the introduction of this measure leads to a number of helpful insights. Specifically, it reveals how a natural assumption of conditional independence for measurement results for conjugate observables leads to a direct demonstration of nonlocality for entangled states, even in cases where the Bell/CHSH inequality is not violated; it shows how, contrary to common belief, a quantum measurement can provide some information, in a clearly defined and quantifiable sense, about the outcome of an unperformed measurement of an observable conjugate to the measured one, and that violation of a Bell inequality can be interpreted as a signature that this "information leakage" is unavoidable. We conclude by considering an alternative perspective on our results, which suggests that quantum mechanics may be able to inform us as to the correct way to reason counterfactually about the physical world.

Definition and classical examples

In order to convey what we are trying to capture with our proposed measure, let's start with a couple of simple examples from the realm of purely classical probability theory. Suppose that we have a collection of *N* fair coins, where *N* is arbitrarily large. We label the coins *i*, with $i \in [1, \cdots N]$. Now suppose that we toss each of the coins once, in sequence, using our right hand. We obtain a random sequence of heads (*H*) and tails (*T*) outcomes, which might for example be *HTTTHHTHHHTHTHTTHTTHHH……….* We will call this set of outcomes the measured results set $\{M_i\}$, and assign each of the heads outcomes the value +1 and each tails outcome the value -1, so that for the set of outcomes just given $M_1 = +1$, $M_2 = M_3 = M_4 = -1$, etc.

Now suppose we try to imagine what might have happened if we had carried out the same exercise, but using our left hand to toss the coins instead of our right. If we had done this, we would have obtained another set of outcomes, the unmeasured results set, which we will call $\{U_i\}$. Now we will of course have no idea what this sequence of results would have been, other than that it would have been another random sequence of heads and tails. However, what we *can* meaningfully attempt to quantify is the correlation between the measured and unmeasured results sets. To see this, consider the standard Pearson correlation coefficient for the two results sets, which we will call $\rho_{MU}$:

$$\rho_{MU} = \frac{E[(M - \mu_M)(U - \mu_U)]}{\sigma_M \sigma_U}$$

where the $\mu$ s and $\sigma$ s denote expectation values and standard deviations respectively. For fair coins, $\mu_M = \mu_U = 0$ and $\sigma_M = \sigma_U = 1$ and so the expression reduces to

$$\rho_{MU} = E(MU) = \underset{N \to \infty}{Lim}\left[\frac{1}{N}\sum_{i=1}^{N} M_i U_i\right].$$

It is not difficult to see that, in this example, $\rho_{MU} = 0$, because we expect there to be an equal number of matches and mismatches between the constituents of each pair of outcomes $(M_i, U_i)$, with each match contributing +1 and each mismatch -1 to the sum.

An objection might be raised at this point, to the effect that quantities such as $\mu_U$, $\sigma_U$, and $\rho_{MU}$ are meaningless, because they refer to counterfactuals – distributions which by definition can never be observed. However, it is perfectly logical to construct a counterfactual example according to simple assumptions that can be agreed on – such as that the coins are fair and that the coin tosser is unable to influence the outcome of each toss. Given these assumptions it is perfectly intuitive (and, arguably, obvious) that the correlation between the measured and unmeasured quantities must equal zero in this example. It should also be remembered that counterfactuals constitute an unavoidable component of derivations of the Bell and CHSH and related inequalities, and have more generally been incorporated into theories of causation ever since Hume used the concept [6], as well as being routinely used for applied statistical analysis in practical fields such as epidemiological analysis [7].

Now let's consider a second example that is still in the classical domain but, in that it involves two different "observables", provides a closer analogy to the quantum mechanical examples that we will come to presently. Suppose that we have a box that contains a large number of objects, half of which are cube shaped and the other half sphere shaped. Of the cubes, three quarters are coloured red and the other quarter blue, whereas three quarters of the spheres are blue and the other quarter red. It follows that, overall, half of the objects are red and the other half blue. Now suppose that we carry out a series of "measurements" that consist of selecting an object from the box at random, using a pair of forceps to remove it, and determining its shape whilst remaining ignorant of its colour. (For example, we could put on a blindfold and then just feel the shape of each selected object.) Following each measurement, the object is replaced in the box. The result of each measurement is recorded, assigning the value +1 for a cube and -1 for a sphere, so we will end up with a random sequence of +1s and -1s, as in the coin-tossing example. We now consider the counterfactual case, where we ask what would have happened if, following the forceps-assisted removal of each object as previously, we had determined its colour but not its shape. (This could be achieved by donning a special pair of glasses that distort the image of each object such that its shape is unrecognisable, whilst its colour remains visible). Suppose a red object is assigned the value +1 and a blue object -1. If we now compute the measured/unmeasured correlation value $\rho_{MU}$ as in the coin tossing case, we find that in this case it is equal to 0.5, and not 0 as in the coin-tossing case. This example is of course extremely contrived, because our prohibition on determining the shape and colour of each object simultaneously is entirely self-imposed. However, it is a useful stepping stone to the quantum mechanical example we consider next, for which the corresponding prohibition on simultaneous determination of our two observables is fundamental.

Quantum mechanical example

We'll now set up a simple example in the quantum mechanical domain. Instead of coins or coloured shapes, we will consider a large ensemble of *N* spin-1/2 particles. We would like the probability for an "up" or "down" result for a spin-component measurement to both be equal to 0.5, regardless of the direction that is chosen for the component measurement; this criterion can be achieved by preparing the particles in the maximally mixed state $(\psi_{mix})_i = \frac{1}{2}\left(|\uparrow\rangle_{i\,i}\langle\uparrow| + |\downarrow\rangle_{i\,i}\langle\downarrow|\right)$ (where once again we are using the label *i* to identify each particle in the ensemble.)

Now suppose we measure the spin component of each particle along a given direction $\underline{a}$. We will again obtain a random sequence of outcomes such as Up, Down, Up, Up, Down, Up….. ; we assign the value +1 to each "up" result and -1 to each "down" result, and we label the full results set as $\{\widetilde{M}_i\}$. We then consider what might have happened if, instead of measuring each spin component along $\underline{a}$ we had measured along the direction $\underline{a}'$, where $\underline{a}.\underline{a}' = 0$. We'll call this unmeasured results set $\{\widetilde{U}_i\}$. As in the coin-tossing case, although we can infer that the results in $\{\widetilde{U}_i\}$ would have consisted of another random sequence of "up" and "down" outcomes, we have no idea what the specific sequence would have been. However, we can once again attempt to evaluate the correlation, $\rho_{\widetilde{M}\widetilde{U}}$, between $\{\widetilde{M}_i\}$ and $\{\widetilde{U}_i\}$. As in the classical coin-tossing example, we find that $\mu_{\widetilde{M}} = \mu_{\widetilde{U}} = 0$ and $\rho_{\widetilde{M}} = \rho_{\widetilde{U}} = 1$, so $\rho_{\widetilde{M}\widetilde{U}} = \underset{N\to\infty}{Lim}\left[\frac{1}{N}\sum_{i=1}^{N}\widetilde{M}_i\widetilde{U}_i\right]$.

Since the measurement directions $\underline{a}$ and $\underline{a}'$ are orthogonal, it follows that, if we were to carry out the $\hat{\sigma}_{\underline{a}}$ and $\hat{\sigma}_{\underline{a}'}$ measurements in succession on our ensemble of particles, the result sets would be completely uncorrelated. One might expect that, likewise, the measured and unmeasured result sets for $\hat{\sigma}_{\underline{a}}$ and $\hat{\sigma}_{\underline{a}'}$ would be completely uncorrelated and hence that $\rho_{\widetilde{M}\widetilde{U}}$ must necessarily be equal to zero, as in the classical coin tossing example. But it would be rash to jump to such a conclusion; we must first consider how our state $(\psi_{nn})_i$ may have been prepared.

Preparation of the maximally mixed state

The two most obvious methods for preparing $\psi_{mix}$ are: (1) compiling an ensemble of randomly oriented spin-1/2 states, or (2) preparing an ensemble of pairs of spin-1/2 particles in the EPR state $\frac{1}{\sqrt{2}}\left(\left|\uparrow_1\downarrow_2\right\rangle-\left|\downarrow_1\uparrow_2\right\rangle\right)$, in which case the marginal state of each of the subsystems will automatically be the maximally mixed state. The multiplicity of realizations of $\psi_{mix}$ can be seen by considering what happens if, in case (1) we carry out a spin-component measurement $\hat{\sigma}_{\underline{\theta}}$, for each member of the ensemble, along some direction $\underline{\theta}$, and then throw away the results of the measurements, or in case (2) we carry out a spin-component measurement, $\hat{\sigma}_{2\underline{\theta}}$ along $\underline{\theta}$ on subsystem 2, and then consider the state of subsystem 1, assuming that the latter is specified based on the knowledge that the $\hat{\sigma}_{2\underline{\theta}}$ measurement has taken place, but without knowing the outcomes of this measurement. In both cases, knowing that a $\hat{\sigma}_{\underline{\theta}}$ (or $\hat{\sigma}_{2\underline{\theta}}$) measurement has taken place will not change our state description – it will remain as $\psi_{mix}=\frac{1}{2}\left(\left|\uparrow\right\rangle\left\langle\uparrow\right|+\left|\downarrow\right\rangle\left\langle\downarrow\right|\right)$.

We can see that the single particle and EPR-pair methods of preparation as described above are equivalent. The EPR-pair method can be interpreted as a remote preparation of our local spin-1/2 particle, and – given that we are assuming for now that the measurement of $\hat{\sigma}_{2\underline{\theta}}$ has taken place in the timelike past with respect to the time at which we specify the state of particle 1 - the effect is identical to a local preparation. We will focus on the EPR-pair approach from now on, as this will enable to link straightforwardly to the Bell/CHSH set-up, which we will come on to later.

<u>Effect of preparation on correlation measure</u>

Suppose that we set $\underline{\theta}$ equal to $\underline{b}$ where $\underline{b}$ lies in the plane of $\underline{a}$ and $\underline{a'}$ and bisects the right angle between them. We will also, for future reference specify another direction $\underline{b'}$, also in the plane of $\underline{a}$ and $\underline{a'}$, where $\underline{b'}$ is at right angles to $\underline{b}$ such that the angle between $\underline{a'}$ and $\underline{b'}$ is 45 degrees and the angle between $\underline{a}$ and $\underline{b'}$ is 135 degrees. Hence $\underline{a}.\underline{b}=\underline{a}'.\underline{b}=\underline{a}'.\underline{b}'=\frac{1}{\sqrt{2}}$, and $\underline{a}.\underline{b}'=-\frac{1}{\sqrt{2}}$. Readers familiar with the CHSH paper will recognize $\underline{a}$, $\underline{a'}$, $\underline{b}$, and $\underline{b'}$ as the apparatus settings that are used in the CHSH set-up. (Because the CHSH experiment used entangled polarizations rather than spins, the angles used by CHSH are half of ours, although all of the algebraic results are identical.)

We will now consider what happens to our local correlation $\rho_{\tilde{M}\tilde{U}}$ between measured and unmeasured quantities, which in this case we can also write as $\rho_{1\underline{a}1\underline{a}'}$. First let's suppose that no measurement at all is carried out on the remote particle (particle 2). In this case there is no reason to expect any correlation between the measured $\hat{\sigma}_{1\underline{a}}$ outcomes and the unmeasured $\hat{\sigma}_{1\underline{a}'}$ outcomes. Indeed, we would be surprised if they were not completely uncorrelated, given that this would definitely be the case if the measurements were carried out consecutively. Let's assume then that, in the absence of any measurement on particle 2, $\rho_{1\underline{a}1\underline{a}'}$ is equal to zero. We can write this as $\rho_{1\underline{a}1\underline{a}'}\big|\neg\hat{\sigma}_2 = 0$.

Now let's consider an alternative scenario, where we know that a measurement of $\hat{\sigma}_{2\underline{b}}$ has been already been carried out, in the timelike past. Can we claim that $\rho_{1\underline{a}1\underline{a}'}$ remains equal to zero in this scenario, i.e. that $\rho_{1\underline{a}1\underline{a}'}\big|\hat{\sigma}_{2\underline{b}} = 0$ ? No, it seems that we can't, as is revealed by a quick calculation. The proportion $P_{2\underline{b}1\underline{a}}$ of the preparatory $\hat{\sigma}_{2\underline{b}}$ results that match the realized $\hat{\sigma}_{1\underline{a}}$ results is (according to quantum mechanics) equal to $\frac{1}{2}\left(1+\frac{1}{\sqrt{2}}\right)$ , and likewise the proportion $P_{2\underline{b}1\underline{a}'}$ of the $\hat{\sigma}_{2\underline{b}}$ results that match the unrealized $\hat{\sigma}_{1\underline{a}'}$ outcomes is also predicted by quantum mechanics to be equal to $\frac{1}{2}\left(1+\frac{1}{\sqrt{2}}\right)$. If we make the apparently natural assumption that the preparatory $\hat{\sigma}_{2\underline{b}}$ outcomes are unaffected by our choice of whether to measure $\hat{\sigma}_{1\underline{a}}$ or $\hat{\sigma}_{1\underline{a}'}$ - which would appear to be obvious (though we will return to this point later) given that the $\hat{\sigma}_{2\underline{b}}$ outcomes occur in the timelike past of the subsequent $\hat{\sigma}_{1\underline{a}}$ or $\hat{\sigma}_{1\underline{a}'}$ measurement – then it follows that the proportion $P_{1\underline{a}1\underline{a}'}$ of the realized $\hat{\sigma}_{1\underline{a}}$ outcomes that match the unrealized $\hat{\sigma}_{1\underline{a}'}$ outcomes must be at least equal to

$P_{2\underline{b}1\underline{a}}$ + $P_{2\underline{b}1\underline{a}'}$ - 1. Hence $P_{1\underline{a}1\underline{a}'} \geq \frac{1}{\sqrt{2}}$, and consequently $\rho_{1\underline{a}1\underline{a}'}\big|\hat{\sigma}_{2\underline{b}} \geq \frac{1}{\sqrt{2}} - \left(1-\frac{1}{\sqrt{2}}\right) = \sqrt{2}-1$.

What is this telling us? The *minimum* correlation between the measured and unmeasured quantities is, in this example, equal to $\sqrt{2}-1$. But to obtain this minimum, there would have to be an implausible level of collusion between the $\hat{\sigma}_{2\underline{b}}$, $\hat{\sigma}_{1\underline{a}}$ and $\hat{\sigma}_{1\underline{a}'}$ outcomes. Specifically, obtaining this minimum correlation of $\sqrt{2}-1$ would require that every single $\hat{\sigma}_{2\underline{b}}$ outcome that did not match the $\hat{\sigma}_{1\underline{a}}$ outcome would have to match the $\hat{\sigma}_{1\underline{a}'}$ outcome, and that, similarly, every single $\hat{\sigma}_{2\underline{b}}$ outcome that did not match the $\hat{\sigma}_{1\underline{a}'}$ outcome would have to match the $\hat{\sigma}_{1\underline{a}}$ outcome.

<u>Conditional independence assumption</u>

A much more natural assumption for the relationship between the measured $\hat{\sigma}_{1\underline{a}}$ and unmeasured $\hat{\sigma}_{1\underline{a}'}$ outcomes is that they should be *conditionally independent*. In other words, we could assume that any correlation between them is entirely a reflection of their level of correlation with the prior $\hat{\sigma}_2$ outcomes. If we follow this assumption, and no $\hat{\sigma}_2$ measurement is performed, then the relation $\rho_{1\underline{a}1\underline{a}'}|\neg\hat{\sigma}_2 = 0$ follows immediately. If on the other hand a $\hat{\sigma}_{2\underline{b}}$ measurement is carried out, then the three correlation pairs between the three pairs of outcomes would display a product structure under the conditional independence assumption. Hence, using the fact that $\rho_{2\underline{b}1\underline{a}} = P_{2\underline{b}1\underline{a}} - (1 - P_{2\underline{b}1\underline{a}}) = \frac{1}{\sqrt{2}}$ (and likewise for $\rho_{2\underline{b}1\underline{a}'}$) we would have:

$$\rho_{1\underline{a}1\underline{a}'}|\hat{\sigma}_{2\underline{b}} = \rho_{2\underline{b}1\underline{a}}\rho_{2\underline{b}1\underline{a}'} = \frac{1}{2}.$$

<u>Implications of non-zero local correlation</u>

It is worth pausing here to consider what this result may be telling us. A non-zero correlation value for $\rho_{1\underline{a}1\underline{a}'}$ implies that every outcome for a $\hat{\sigma}_{1\underline{a}}$ measurement reveals a finite amount of information about what the result of a $\hat{\sigma}_{1\underline{a}'}$ measurement would have been, even though $\hat{\sigma}_{1\underline{a}}$ and $\hat{\sigma}_{1\underline{a}'}$ are complementary observables. In order to make such an inference, we need to know the specific $\hat{\sigma}_2$ measurement that has previously taken place, but we don't need to know the outcome of this measurement. Under these conditions, we can make quantitative statements with regard to the amount of information pertaining to an unmeasured $\hat{\sigma}_{1\underline{a}'}$ outcome that is revealed by each $\hat{\sigma}_{1\underline{a}}$ outcome. This can be represented as:

$$Info\left(\hat{\sigma}_{1\underline{a}'}\middle|\hat{\sigma}_{1\underline{a}},\underline{\theta}\right) = 1 - H\left(P_{1\underline{a}1\underline{a}'}\middle|\underline{\theta}\right)$$

where $H(\ )$ is the Shannon entropy function, $H(x) = -(x\ln_2 x + (1-x)\ln_2(1-x))$, and $\underline{\theta}$ parameterizes the $\hat{\sigma}_2$ measurement that has taken place, i.e. $\hat{\sigma}_2 \to \sigma_{2\underline{\theta}}$.

Given that $\hat{\sigma}_{1\underline{a}}$ and $\hat{\sigma}_{1\underline{a}'}$ are complementary, and maintaining our conditional independence assumption for the $\hat{\sigma}_{1\underline{a}}$ and $\hat{\sigma}_{1\underline{a}'}$ outcomes, we find that $Info\left(\hat{\sigma}_{1\underline{a}'}\middle|\hat{\sigma}_{1\underline{a}},\underline{\theta}\right)$ is maximized when $\underline{\theta} = \underline{b}$ where $\underline{b}$ is defined exactly as in the CHSH-type set-up, as described above. For example, if instead we were to set $\underline{\theta} = \underline{c}$, where $\underline{c}$ is another direction coplanar with $\underline{a}$ and $\underline{a}'$ we find that $\rho_{1\underline{a}1\underline{a}'}|\underline{c} = \underline{a}.\underline{c}\left(\sqrt{1-(\underline{a}.\underline{c})^2}\right) < \rho_{1\underline{a}1\underline{a}'}|\underline{b}$. When $\underline{\theta} = \underline{b}$, $\rho_{1\underline{a}1\underline{a}'} = 0.5$, from which it follows that $P_{1\underline{a}1\underline{a}'} = 0.75$. Hence,

$$Max\left[Info\left(\hat{\sigma}_{1\underline{a}'}\middle|\hat{\sigma}_{1\underline{a}},\underline{\theta}\right)\right] = 1 - H(0.75) \approx 0.19 \text{ bits.}$$

This means that when we carry out a $\hat{\sigma}_{1\underline{a}}$ measurement under these conditions, then each outcome provides a total of about 1.19 bits, made up of 1 bit from the $\hat{\sigma}_{1\underline{a}}$ outcome, and about 0.19 additional bits from what we know about the unperformed $\hat{\sigma}_{1\underline{a}'}$ outcome. This figure of 1.19 bits is the same as the amount of classical resources that Cerf, Gisin, and Massar [8] found were required to simulate an EPR-pair, though this is probably just a coincidence.

<u>Spacelike separated measurements and nonlocality implications</u>

Up to this point we have restricted the discussion to cases where the $\hat{\sigma}_2$ measurements occur in the timelike past of the $\hat{\sigma}_1$ measurements. Whilst this simplifies the calculation of $\rho_{1\underline{a}1\underline{a}'}$ because it provides a natural justification for assuming that the $\hat{\sigma}_2$ outcomes are unaffected by our choice of $\hat{\sigma}_1$ measurement, it doesn't enable us to address the issue of nonlocality, which is the purpose of Bell/ CHSH experiments. In order to fully replicate the CHSH set-up, we need to stipulate that our $\hat{\sigma}_2$ measurement events are spacelike separated with respect to the $\hat{\sigma}_1$ measurements. In this case the $\hat{\sigma}_2$ outcomes may depend on our choice of $\hat{\sigma}_1$ measurement – but only if nonlocal effects are at work. So, if our aim is to determine whether nonlocality is necessarily implicated, we can start by maintaining the assumption that the $\hat{\sigma}_2$ outcomes are unaffected by the choice of $\hat{\sigma}_1$ measurement, and see what the implications are for possible nonlocality effects in the other direction, consistent with this assumption.

With this in mind, we make the following bold claim:

- If the value of $\rho_{1\underline{a}1\underline{a}'}$ *necessarily* depends on whether or not a $\hat{\sigma}_2$ measurement is carried out, then an inference of nonlocality is unavoidable.

We justify this claim on the basis that a change in the implied $\rho_{1\underline{a}1\underline{a}'}$ can be interpreted as a real physical change, even though it cannot be directly observed. In a classical context, such an interpretation would be perfectly natural: in the coloured shapes example we considered earlier, the corresponding change in implied correlation would equate to a change in the relative proportions of spheres and cubes that were red as opposed to blue.

Let's now examine the consequences of this claim, within our framework, and how they compare with the usual conclusions with regard to the CHSH set-up. The CHSH inequality is normally written as:

$$S = \left| E(\underline{a},\underline{b},) - E(\underline{a},\underline{b}') + E(\underline{a}',\underline{b}) + E(\underline{a}',\underline{b}') \right| \leq 2,$$

where, according to our notation, $E(\underline{\alpha},\underline{\beta}) = \rho_{2\underline{\beta}1\underline{\alpha}}$. With the parameter values as we have set them (i.e. $\underline{a}.\underline{b} = \underline{a}'.\underline{b} = \underline{a}'.\underline{b}' = \frac{1}{\sqrt{2}}$, $\underline{a}.\underline{b}' = -\frac{1}{\sqrt{2}}$), *S* attains its maximum possible value (according to quantum mechanics) of $2\sqrt{2}$. (This is known as Tsirelson's bound [9].)

We now consider a scenario in which the inequality is not violated, whilst still retaining the condition that $\underline{a}.\underline{a}'=0$. This can be achieved by leaving $\underline{a}$, $\underline{a}'$, and $\underline{b}$ unchanged, and setting $\underline{b}'$ such that it is co-planar with $\underline{a}$ and $\underline{a}'$ and makes an angle of 55 degrees with $\underline{a}$ and 35 degrees with $\underline{a}'$. With these settings we find that the CHSH expression $S \approx 1.442$, and so the inequality is not violated. Given that $\underline{a}.\underline{a}'=0$, we once again have the commutation relation $\left[\hat{\sigma}_{1\underline{a}}, \hat{\sigma}_{1\underline{a}'}\right] = i\hbar$, and hence it is once again natural to assume that the conditional independence assumption should apply with respect to the $\hat{\sigma}_{1\underline{a}}$ and unmeasured $\hat{\sigma}_{1\underline{a}'}$ outcomes. Following this assumption, $\rho_{1\underline{a}1\underline{a}'} = 0$ if no measurement is carried out on particle 2. If a measurement of $\hat{\sigma}_{2\underline{b}}$ is carried out, $\rho_{1\underline{a}1\underline{a}'} = \rho_{2\underline{b}1\underline{a}}\rho_{2\underline{b}1\underline{a}'} = E(\underline{a},\underline{b})E(\underline{a}',\underline{b}) = 0.5$. If, instead, a measurement of $\hat{\sigma}_{2\underline{b}'}$ is carried out, $\rho_{1\underline{a}1\underline{a}'} = \rho_{2\underline{b}'1\underline{a}}\rho_{2\underline{b}'1\underline{a}'} = E(\underline{a},\underline{b}')E(\underline{a}',\underline{b}') = 0.47$. So, in both of the cases where a measurement is carried out on particle 2, we find that $\rho_{1\underline{a}1\underline{a}'} = 0$ changes and this change can take effect over a spacelike surface. So, following our earlier claim, nonlocality is unavoidably present, even though the CHSH inequality is not violated. It is interesting that the demonstration of nonlocality in cases such as this trumps the non-violation of the CHSH inequality with the given parameter settings.

Measured/ unmeasured correlation in the no-signalling boxes framework

The framework used by Pawlowski et al in their recent pioneering work on Information Causality [2] – in which they provided a principle underpinning Tsirelson's bound – made use of the concept of no-signalling boxes (NS boxes), which are a generalization of the well-known "PR box" (named after its originators [10]). Whereas the PR box models ultra-quantum no-signalling correlations that transcend Tsirelson's bound and enable the CHSH expression to reach its maximum possible value of 4, the NS boxes framework enables a representation of the full range of possible no-signalling correlations, within a context that is not theory specific. Using the notation of [2], NS boxes are parameterized in terms of probabilities that represent the correlations between inputs *a*, *b*, and outputs *A*, *B* of two spatially separated boxes. The local outputs of the boxes are uniformly random. Within this framework, the CHSH expression can be re-written as

$S = \left|\sum_{a,b} P(A \oplus B = ab | a,b)\right|$ where the inputs are 0 or 1, *P*() indicates probability, and $\oplus$ is addition modulo 2. (In the CHSH spin-measurement set-up we have used, *a* and *b* now correspond to the choice of apparatus settings at the two locations: $\underline{a} \rightarrow a = 0$, $\underline{a}' \rightarrow a = 1$, $\underline{b} \rightarrow b = 0$, and $\underline{b}' \rightarrow b = 1$. *A* and *B* can each take the value 0 or 1, according to whether the outcome "up" or "down" is obtained for a spin-component measurement at the respective locations.)

Since the boxes are no-signalling, it is essential that the correlation between the outcomes of *consecutive* local measurements, with different apparatus settings, at one of the two locations is not dependent on whether or not a measurement is carried out at the remote location, or on the remote apparatus setting if the remote measurement is performed. (If this were not the case it would be possible to signal nonlocally.) However, when a remote measurement is performed, the correlation between measured and unmeasured local quantities – in the sense we have discussed, where only one of the measurements is performed and the other is considered in a counterfactual sense – will be subject to constraints which will not, in general, permit equality between (i) the correlation between the measurement outcomes when considered as actual consecutive measurements, and (ii) the correlation between the outcomes where only one measurement is performed and the other is considered in a counterfactual sense. And since we are using correlation (i) as an indication of what correlation (ii) is when a remote measurement is *not* performed, the implication is that the remote measurement must modify correlation (ii) and hence must have a nonlocal influence.

For example, suppose that we assume that correlation (i) between actually performed local measurements is zero (as in the CHSH set-up) and that we set $b$ = 0. Then we find that, if we assume that the value of B obtained with this setting is independent of the choice for the $a$ setting, then the proportion of A outcomes obtained with $a$ = 0 that would have given the same outcome had we set $a$ = 1 must be equal to at least $(P_{\min}|b=0) = P(A=B|a=0,b=0)+P(A=B|a=1,b=0)-1$. (We have used here the fact that $P(A=B)=P(A\oplus B=0)$.) Hence the correlation $\rho_{a=0,a=1}$ between the measured and unmeasured $A$ outcomes must be equal to at least $(\rho_{\min}|b=0) = P_{\min}-(1-P_{\min}) = 2[P(A=B|a=0,b=0)+P(A=B|a=1,b=0)]-3$.

Using the approach of [2], we can further simplify the NS box model by assuming all of the $P$()s in the CHSH expression are equal, i.e. we can adopt the assumption that

$$P(A\oplus B=0|a=0,b=0)=P(A\oplus B=0|a=0,b=1)=P(A\oplus B=0|a=1,b=0)=P(A\oplus B=1|a=1,b=1)$$

With this assumption we can see that the condition for $\rho_{\min}$ to be > 0 is given by $P(A\oplus B=ab|a,b)>0.75$. Interestingly, this is also precisely the necessary and sufficient condition for violation of the CHSH inequality. Thus $\rho_{\min}>0$ could be interpreted as a signature of nonlocality, just as violation of the CHSH inequality typically is. Intuitively, this seems natural, as the condition $\rho_{\min}>0$ implies that performing a measurement at the $b$ location enforces a positive correlation between measured and unmeasured $A$ outcomes, whereas otherwise we might have expected these quantities to be uncorrelated.

If on the other hand we also assume conditional independence for the measured and unmeasured outcomes – i.e. if we assume that $\rho_{a=0,a=1}|b=\rho_{a=0,b}\rho_{a=1,b}$, then we find that $\rho_{a=0,a=1}$ is positive in almost all cases of NS boxes. In fact, the conditional independence

assumption implies that $\rho_{a=0,a=1}|b = [2P(A \oplus B = ab|a,b) - 1]^2$ , and so it follows that the only non-positive value for $\rho_{a=0,a=1}$ occurs when $P(A \oplus B = ab|a,b) = 0.5$, in which case $\rho_{a=0,a=1} = 0$. But in this special case we also have $\rho_{a,b} = 0$ which means that the $A$ and $B$ outcomes are completely independent and uncorrelated. In all other cases, $\rho_{a=0,a=1} > 0$, whether or not the CHSH inequality is violated. This means that, apart from the trivial exception that we have identified, *all* NS boxes have intrinsic nonlocality if we adopt the (very natural) conditional independence assumption.

(If the correlation between actually performed local measurements is non-zero – as it would be, for example, in a modified CHSH set-up where $\underline{a}.\underline{a}' \neq 0$ - then the above argument based on $\rho_{\min}$ may have to be modified, as with certain settings it could be necessary for $\rho_{a=0,a=1}$ to pass some minimum positive value before we can infer nonlocality.)

Re-visiting counterfactual reasoning

We'll close with a few remarks about the use of counterfactual reasoning. Most of our arguments have, among other things, involved the apparently innocuous assumption that it is meaningful to consider an alternative world where a single different free-will choice was made, whilst other things that couldn't possibly have been affected by this free-will choice are held fixed. For example, we considered a case where a measurement has been carried out on a quantum system in the timelike past with respect to the current measurement; we then assumed that the outcome of the measurement in the past could be considered as fixed, whilst considering an alternative, counterfactual, choice of current measurement. This is the sort of assumption that can slip in almost unnoticed in counterfactual arguments. The timelike past – in particular those aspects of it, such as measurement outcomes - that have been captured as classical record – is absolutely indelible, right? If we allow that a free will choice can modify the past, at the macroscopic classical level, where would that leave us?

But things are (of course) not as obvious and straightforward as this. For a start, the way we have phrased these questions takes it for granted that "measurement outcomes", "classical record", and "macroscopic" are clearly delineated concepts. Once we accept that any such delineation is blurry at best, we may inadvertently find ourselves slipping into the pure quantum realm, where the potential for retroactive influences, in the form of delayed choice experiments, zig-zagging spacetime diagrams, etc., is seriously advocated by respected theorists [11]. But, more fundamentally, is it possible that quantum mechanics can inform us about the correct way to apply counterfactual reasoning (rather than vice-versa)?

Most of our analysis has been concerned with spacelike, rather than timelike, separated events, and this is the realm in which discussions about nonlocality are normally conducted. In analyzing CHSH type experiments using our proposed correlation measure, we argued that, starting from an assumption of locality, it was legitimate to take a spacelike separated outcome as fixed and to explore the implications of this. The justification for this was the locality premise. The issue of whether, *aside from any locality considerations*, it is meaningful to model a counterfactual quantum mechanical world where we change one thing – something determined by a free-will choice, like an apparatus setting – whilst holding everything that is apparently unconnected with this choice fixed, was not addressed at all.

But perhaps this is a question that is worthy of being raised. Perhaps the profound holism of quantum mechanics can inform us about matters beyond the realm of prediction in the real world; perhaps it can inform us about how we should use our imagination, how we should re-construct a world where things happened differently to the way they actually occurred. This kind of imaginative re-construction is, of course, something that we do all the time in everyday life. It seems likely that the way we (unconsciously) do this is conditioned by our everyday experiences, which (for most people) don't encompass the trans-classical quantum domain.

We're now straying into areas of philosophy that are way beyond the scope of this article. The aim of these closing remarks is simply to raise the question: is it perhaps misguided to imagine a counterfactual world where the only difference from the actual one is a single (apparently free-willed) event, such as the choice of an apparatus setting? What if the holistic nature of quantum mechanics extends to histories as well as states, and that, consequently, we are not at liberty to tweak one event in a history whilst leaving everything else fixed? Then it would be possible that the measured/unmeasured correlation $\rho_{1\underline{a}1\underline{a'}}$ would always be zero if it relates to complementary observables, even if conditioning on the outcome an earlier preparatory measurement indicates that it must be positive. The implication of this would be that *we cannot assume that the outcome of the preparatory measurement would necessarily have been the same in the counterfactual world, even if it was recorded in the timelike past*.

This may seem like a crazy idea, suggesting that we can change the past, or that there cannot be any free will. But I don't think that it would necessitate either of these things. Rather, it may require us to reconcile the existence of free will with the notion that alternative free will choices can only be considered as part of a whole alternative history – a different filtration of stochastic events, if you like. Such a reconciliation may be challenging, but I don't see why it should be impossible. Perhaps there is an element of a free will choice that is outside of spacetime (a status that, it has recently been suggested, could explain how Nature arranges Bell-type correlations [12]).

By exploring possibilities such as this, as a means to unravel Bell's theorem, we are moving the focus away from locality and on to counterfactuals. Recently it has been argued [3] that the role of counterfactuals in the derivation of the CHSH inequality is less significant than their role in deriving the original Bell inequality: counterfactual "meaningfulness" in CHSH replaces counterfactual definiteness in Bell's original. But no-one denies that CHSH contains counterfactuals, albeit in a different guise. Given that any Bell-type inequality necessarily involves combining results from incompatible experimental set-ups, counterfactuals will always play a role in their derivation.

## Concluding remarks

Most of our investigation here has been concerned with looking at CHSH, and related scenarios, from a new perspective, where we are concerned with correlations between two measurements in the same location, rather than separated ones. Just as in the derivation of Bell, CHSH, and all related inequalities, our analysis involves counterfactuals. We have, for the most part, used counterfactual reasoning in the same way that it is commonly employed, using techniques such as conditioning on well-defined past events, or on events in a spacelike separated region once we have assumed locality. This reasoning leads to some interesting results, shedding some new light on nonlocality, beyond what can be inferred from violation or non-violation of Bell/CHSH inequalities.

Following this analysis, we have raised the possibility that the whole approach to counterfactuals may have to be reconsidered when we are in the quantum domain; the suggestion is that quantum mechanics may itself be able to inform us about what is correct and what is misguided in the realm of counterfactual reasoning. This suggestion is purely speculative: if justified, it could lead to some radical new insights. The price to pay could, however, be that a good deal of prior research – including most of what has been presented here – would become invalid.